\documentclass[multphys,vecphys]{svmult}
\usepackage{makeidx}         
\usepackage{graphicx}        
\usepackage{multicol}        
\usepackage[bottom]{footmisc}
\makeindex             
\begin{document}

\title*{Globular Clusters at the Centre of the Fornax Cluster: 
Tracing Interactions Between Galaxies}
\titlerunning{Globular clusters at the centre of the Fornax Cluster} 
\author{Lilia P. Bassino\inst{1,2}\and Tom Richtler\inst{3}\and   
Favio R. Faifer\inst{1,2}\and Juan C. Forte $^{1}$\and Boris 
Dirsch\inst{3}\and Doug Geisler\inst{3}\and Ylva Schuberth\inst{4}}
\authorrunning{L. P. Bassino et al.}
\institute{Facultad de Ciencias Astron\'omicas y Geof\'{\i}sicas, 
Universidad Nacional de La Plata, Paseo del Bosque S/N, 1900-La Plata, 
Argentina
\texttt{lbassino@fcaglp.unlp.edu.ar}
\and 
IALP -- CONICET, Argentina
\and
Universidad de Concepci\'on, Departamento de F\'{\i}sica,
Casilla 160-C, Concepci\'on, Chile
\and
Argelander-Institut f\"ur Astronomie, Auf dem H\"ugel 71, D-53121 Bonn, 
Germany}
\maketitle

\begin{abstract} 
We present the combined results of two investigations: 
a large-scale study of the globular cluster system (GCS) around NGC\,1399, 
the central galaxy of the Fornax cluster, and a study of the GCSs around 
NGC\,1374, NGC\,1379 and NGC\,1387, three low-luminosity early-type 
galaxies located close to the centre of the same cluster. 
In both cases, the data consist of images from the wide-field 
MOSAIC Imager of the CTIO 4--m telescope, obtained with Washington
$C$ and Kron--Cousins $R$ filters, which provide good metallicity 
resolution. 
 
The colour distributions and radial projected densities of the 
GCSs are analyzed.	
We focus on the properties of the GCSs that trace possible interaction 
processes between the galaxies, such as tidal stripping of globular 
clusters (GCs). For the blue GCs, we find tails between NGC\,1399 and 
neighbouring galaxies in the azimuthal projected distribution, and 
the three low-luminosity galaxies show low specific frequencies and 
a low proportion of blue GCs.

\end{abstract}

\section{Introduction}
It is widely known that GCs are a useful tool for studying the origin 
and evolution of galaxies, in the case of isolated galaxies as well as 
for those within groups or clusters.

In our first wide-field CCD study of the GCS around NGC\,1399, at the 
centre of the Fornax cluster \cite{dir03}, we found that there are GCs 
out to the very limits of the studied field (Field 3 in Fig.~1), 
corresponding to a projected galactocentric distance of 100\,kpc. 
In a later run, we obtained images of three adjoining fields, 
using the same observational set up, which is described in Sect.~2. 
We added one field to the West (Field 4 in Fig.~1) and two fields 
to the East of NGC\,1399 (Fields 1 and 2 in Fig.~1). As there were 
several low-luminosity early-type galaxies with their own GCSs in 
the western field (NGC\,1374, NGC\,1379 and NGC\,1387), 
we decided to keep this field to study them \cite{bas06a}, and we used 
the eastern fields, where there were no conspicuous galaxies, to study 
the NGC\,1399 GCS over a larger field \cite{bas06b}. 

\begin{figure}
\centering
\includegraphics[height=8cm]{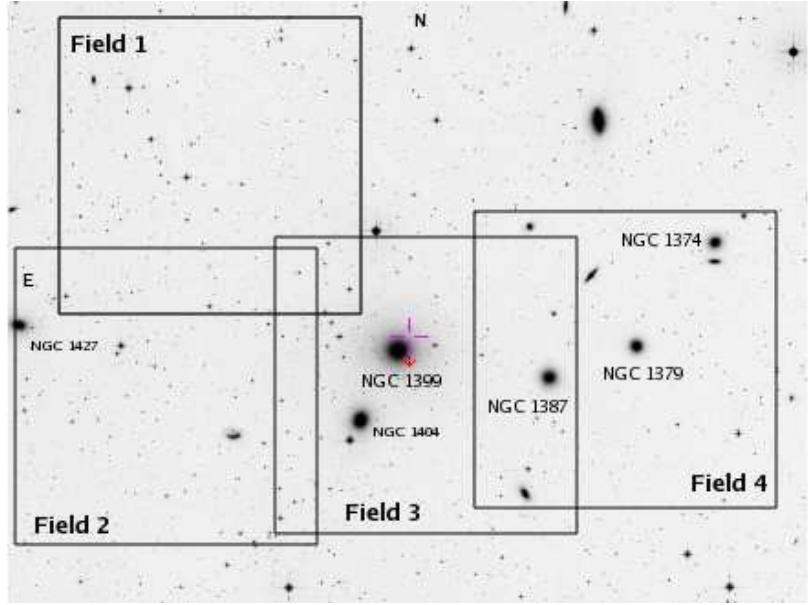}
\caption{MOSAIC fields overlaid on a DSS image of the Fornax cluster. 
North is up and East to the left.}
\end{figure}

\section{Observations and Data Reduction}
The observations were performed with the MOSAIC camera and 4--m Blanco 
telescope at the Cerro Tololo Inter--American Observatory (CTIO).
The MOSAIC wide-field camera has a field of view of 36 $\times$ 36\,arcmin  
(200 $\times$ 200\,kpc at the Fornax distance). For more information on 
the MOSAIC camera we refer to the homepage 
{\it http://www.noao.edu/kpno/mosaic/mosaic.html}.

Kron--Cousins $R$ and Washington $C$ filters were used. We remind the 
reader that $R$ and Washington $T_1$ magnitudes are very similar, 
with a very small colour term and zero-point difference \cite{gei96}. 
For more details on the point source selection, photometric calibration, 
and the identification of GC candidates we refer the reader to 
\cite{bas06a} and \cite{bas06b}. 
Statistical subtraction of the contamination by the background was 
performed in all cases. 

\section{GCSs around NGC\,1374, NGC\,1379 and NGC\,1387}
It is shown, for the first time, that the colour distributions of these  
three low-luminosity galaxies are clearly bimodal, with very similar 
colours for the blue GC peaks \cite{bas06a}. The red peak in NGC\,1387, 
the galaxy located closer to NGC\,1399 (see Fig.~1), 
is redder than the others and its whole colour distribution is atypical: 
the red clusters are much more numerous than the blue ones and the 
separation of the peaks is very pronounced. In fact, the fraction 
of blue clusters, with respect to the total GC population is 
low for the three systems, but even lower for NGC\,1387 (43\%, 45\%, 
and 24\% for NGC\,1374, NGC\,1379 and NGC\,1387, respectively). 
  
With regard to the radial distributions, the blue GCs in these systems 
show flatter distributions than the red ones, while the respective 
galaxy light profiles follow the density profiles for all GCs.

By means of the luminosity functions we estimate the total GC populations 
in NGC\,1374, NGC\,1379 and NGC\,1387 between 200 and 400 clusters, and 
obtain specific frequencies $S_{N}$ = 2.4, 1.4 and 1.8, respectively. 
These specific frequencies are rather small when compared to the 
typical value $S_{N}$~=~4 found for elliptical galaxies in dense 
environments \cite{har03}. 
 
Figure~2 shows the specific frequencies and fractions of blue clusters 
versus distance from NGC\,1399, including the data for NGC\,1404 
\cite{for98} and NGC\,1427 \cite{fort01} (see Fig.~1). The specific 
frequencies seem to decrease with decreasing distance from NGC\,1399 
and a similar trend is present in the fraction of blue clusters. 
These behaviours support the idea that galaxies closer 
to NGC\,1399, in projected distance, might be losing their blue GCs 
as a result of some interaction process, like tidal stripping of 
clusters by the giant elliptical NGC\,1399 (e.g. \cite{for97}, 
\cite{kis99}, \cite{bek03}, etc.).

\begin{figure}
\centering
\includegraphics[height=8cm]{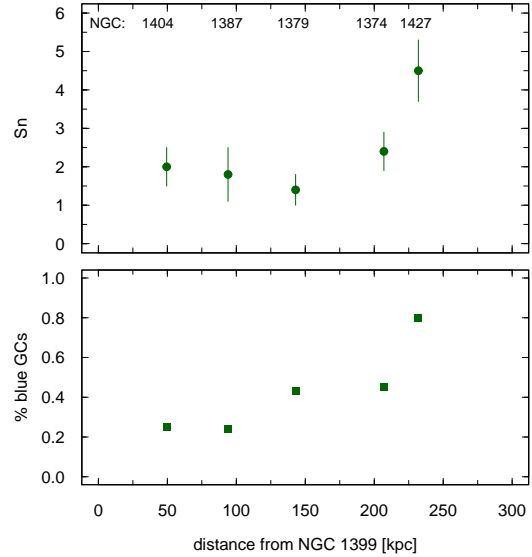}
\caption{Specific frequencies (upper panel) and fraction of blue GCs 
(lower panel) versus projected distance from NGC\,1399.}
\end{figure}

\section{GCS around NGC\,1399} 
The large-scale study of the projected radial distribution of GCs 
around NGC\,1399 \cite{bas06b} shows that the blue clusters extend 
up to 250\,kpc from the galaxy centre (45 $\pm$ 5\,arcmin) while 
the red ones show a steeper radial profile that reaches a radius 
of 140\,kpc (25 $\pm$ 5\,arcmin). 
To our knowledge, this is one of the largest GCSs ever studied and, 
as our limiting magnitude is $T_1 = 23$, we cannot assure that there 
are no more GCs further out.

The colour distributions at different radial ranges confirm that the  
red GCs are more centrally concentrated than the blue ones, and show  
that the blue peak gets bluer with increasing galactocentric 
radius; a similar gradient was found in the NGC\,1427 GCS by 
\cite{fort01}.    	

The colour distributions in different magnitude ranges show that  
the distribution is unimodal for the brightest bin (magnitudes similar  
to that of $\omega$\,Cent), as already pointed out by \cite{dir03}. 
However, we found no evidence for a ``blue tilt" (e.g. \cite{har06}), i.e. 
the blue peak does not get redder with increasing luminosity. 

The azimuthal distribution of the smoothed projected number density 
of blue clusters around NGC\,1399 (Fig.~9 in \cite{bas06b}) 
shows two tails: one towards NGC\,1404 and another towards NGC\,1387. 
In the case of NGC\,1404 one might wonder if this is just an overlapping 
of the two GCSs but it has been proposed, and tested by numerical 
simulations \cite{bek03}, that its GCs are probably being stripped 
by NGC\,1399. The other tail, towards NGC\,1387 cannot be just an 
overlapping of GCSs due to the small size of the GCS of NGC\,1387 
(r = 3\,arcmin \cite{bas06a}) as compared to the projected distance 
to NGC\,1399 (19\,arcmin). Such overdensity of blue GCs may be 
understood as evidence that blue clusters, the less bound ones, 
are being lost by NGC\,1387 due to some interaction process with 
the central cluster galaxy, like tidal stripping.

\printindex

\begin{thebibliography}{99.}
\bibitem{bas06a}Bassino, L.P., Richtler, T., Dirsch, B.: MNRAS \textbf{367}, 156 (2006)
\bibitem{bas06b}Bassino, L.P., Faifer, F.R., Forte, J.C., Dirsch, B., 
Richtler, T., Geisler, D. and Schuberth, Y.: A\&A, in press, 
astro-ph/0603349 (2006)
\bibitem{bek03} Bekki K., Forbes D.A., Beasley M.A., Couch W.J.: MNRAS 
\textbf{344}, 1334 (2003)
\bibitem{dir03}Dirsch B., Richtler T., Geisler D., et al.: AJ \textbf{125}, 
1908 (2003)
\bibitem{for97}Forbes D.A., Brodie J.P., Grillmair C.J.: AJ \textbf{113}, 
1652 (1997)
\bibitem{for98}Forbes D.A., Grillmair C.J., Williger G.M. et al.: 
MNRAS \textbf{293}, 325 (1998)
\bibitem{fort01}Forte J.C., Geisler D., Ostrov P.G. et al.: 
AJ \textbf{121}, 1992 (2001)
\bibitem{gei96}Geisler D.: AJ \textbf{111}, 480 (1996)
\bibitem{har03} Harris W.E.: in Kissler--Patig M., ed., 
Extragalactic Globular Cluster Systems, ESO Astrophysics Symposia, 
Springer--Verlag, Berlin, p. 317 (2003)
\bibitem{har06}Harris, W.E., Whitmore, B.C., Karakla, D. et al.: ApJ 
\textbf{636}, 90 (2006)
\bibitem{kis99}Kissler--Patig M., Grillmair C.J., Meylan G. et al.:
AJ \textbf{117}, 1206 (1999)
\end{thebibliography}
\end{document}